\documentclass[twocolumn,aps,pra,showpacs,superscriptaddress]{revtex4}

\usepackage{graphicx}
\usepackage{amsfonts}
\usepackage{amsmath}
\usepackage{amssymb}
\usepackage[colorlinks=true,citecolor=blue,urlcolor=blue]{hyperref}

\def\endproof{\vrule height6pt width6pt depth0pt}

\begin{document}

%%%%%%%%%%%%%%%%%%%%%%%%%%%%%%%%%%%%%%%%%%%%%%%%%%%%%%%%%%%%%%%%%%%

\title{Thermodynamical cost of some interpretations of quantum theory}

%%%%%%%%%%%%%%%%%%%%%%%%%%%%%%%%%%%%%%%%%%%%%%%%%%%%%%%%%%%%%%%%%%%

% K.Wiesner@bristol.ac.uk,enxkw@bristol.ac.uk,otfried.guehne@uni-siegen.de,ceptryn@gmail.com,jan-ake.larsson@liu.se

%%%%%%%%%%%%%%%%%%%%%%%%%%%%%%%%%%%%%%%%%%%%%%%%%%%%%%%%%%%%%%%%%%%

\author{Ad\'an Cabello}
 \email{adan@us.es}
 \affiliation{Departamento de F\'{\i}sica Aplicada II, Universidad de Sevilla, E-41012 Sevilla, Spain}

\author{Mile Gu}
 \affiliation{School of Physical and Mathematical Sciences, Nanyang Technological University, 21 Nanyang Link, Singapore 637371, Singapore}
 \affiliation{Complexity Institute, Nanyang Technological University, 18 Nanyang Drive, Singapore 637723, Singapore}
 \affiliation{Centre for Quantum Technologies, National University of Singapore, 3 Science Drive 2, Singapore 117543, Singapore}

\author{Otfried G\"uhne}
 \affiliation{Naturwissenschaftlich-Technische Fakult\"at, Universit\"at Siegen,
 Walter-Flex-Stra{\ss}e 3, D-57068 Siegen, Germany}

\author{Jan-{\AA}ke Larsson}
 \affiliation{Institutionen f\"or Systemteknik, Link\"opings Universitet, SE-58183 Link\"oping, Sweden}

\author{Karoline Wiesner}
 \affiliation{School of Mathematics, Centre for Complexity Sciences, University of Bristol, University Walk, Bristol BS8 1TW, United Kingdom}

%%%%%%%%%%%%%%%%%%%%%%%%%%%%%%%%%%%%%%%%%%%%%%%%%%%%%%%%%%%%%%%%%%%

\date{\today}

%First version: September 23, 2014 (Siegen).
%This version: May 4, 2016 (Sevilla).

%%%%%%%%%%%%%%%%%%%%%%%%%%%%%%%%%%%%%%%%%%%%%%%%%%%%%%%%%%%%%%%%%%%

\begin{abstract}
The interpretation of quantum theory is one of the longest-standing debates in physics. Type~I interpretations see quantum probabilities as determined by intrinsic properties of the observed system. Type~II see them as relational experiences between an observer and the system. It is usually believed that a decision between these two options cannot be made simply on purely physical grounds but requires an act of metaphysical judgment. Here we show that, under some assumptions, the problem is decidable using thermodynamics. We prove that type~I interpretations are incompatible with the following assumptions: (i) The choice of which measurement is performed can be made randomly and independently of the system under observation, (ii) the system has limited memory, and (iii) Landauer's erasure principle holds.
\end{abstract}

%%%%%%%%%%%%%%%%%%%%%%%%%%%%%%%%%%%%%%%%%%%%%%%%%%%%%%%%%%%%%%%%%%%

%03.65.Ta: Foundations of quantum mechanics; measurement theory
%03.65.Ud: Entanglement and quantum nonlocality
%(e.g. EPR paradox, Bell's inequalities, GHZ states, etc.)

\maketitle

%%%%%%%%%%%%%%%%%%%%%%%%%%%%%%%%%%%%%%%%%%%%%%%%%%%%%%%%%%%%%%%%%%%

\section{Introduction}

%%%%%%%%%%%%%%%%%%%%%%%%%%%%%%%%%%%%%%%%%%%%%%%%%%%%%%%%%%%%%%%%%%%

The interpretations of quantum theory can be classified \cite{Cabello15} in two types: Those that view quantum probabilities of measurement outcomes as determined by intrinsic (observer-independent) properties of the observed system, for example, Einstein's \cite{Einstein36}, Bohmian mechanics \cite{Bohm52,Goldstein13}, many worlds \cite{Everett57,Vaidman15}, Ballentine's \cite{Ballentine70}, modal interpretations \cite{vanFraassen72,LD12}, Bell's beables \cite{Bell76}, collapse theories \cite{GRW86,Ghirardi11}, and Spekkens's \cite{Spekkens07}, and those according to which quantum theory does not deal with intrinsic properties of the observed system but with the experiences an observer has of the observed system, for example, Copenhagen \cite{Bohr98,Faye14}, Wheeler's \cite{Wheeler83}, relational \cite{Kochen85,Rovelli96}, Zeilinger's \cite{Zeilinger99}, Fuchs and Peres's no interpretation \cite{FP00}, and QBism \cite{Fuchs10,FMS14}. Here, following Ref.\ \cite{Cabello15}, we call them type I and type II interpretations, respectively. It is usually believed that deciding between these two options ``cannot be made simply on purely physical grounds but it requires an act of metaphysical judgement'' \cite{Polkinghorne14}.

While the actual classification of interpretations itself is an active and interesting subject of discussion, the aim of this article is to show that type I interpretations are incompatible with the following assumptions: (i) The choice of which measurement is performed on a system under observation can be made randomly and independently of the quantum system, (ii) the system has limited memory, and (iii) Landauer's erasure principle \cite{Landauer61} holds. The article is organized as follows. In Sec.\ \ref{Sec:2} we introduce the ideal experiment we will consider in the rest of the article and list our assumptions. In Sec.\ \ref{Sec:3} we introduce the optimal finite-state machine needed for rigorously proving our results. The results and their proofs are presented in Sec.\ \ref{Sec:4}. Finally, in Sec.\ \ref{Sec:5} we summarize our conclusions.

%%%%%%%%%%%%%%%%%%%%%%%%%%%%%%%%%%%%%%%%%%%%%%%%%%%%%%%%%%%%%%%%%%%

\section{Scenario}
\label{Sec:2}

%%%%%%%%%%%%%%%%%%%%%%%%%%%%%%%%%%%%%%%%%%%%%%%%%%%%%%%%%%%%%%%%%%%

Consider the following ideal experiment: A single qubit is sequentially measured at time intervals by an observer who performs projective measurements randomly chosen between the Pauli observables $\sigma_z$ and $\sigma_x$. Each of these measurements has two possible outcomes: $+1$ or $-1$. When measuring $\sigma_z$, if the outcome is $+1$, the quantum state after the measurement is $|0\rangle$; if the outcome is $-1$, the quantum state is $|1\rangle$. Similarly, when measuring $\sigma_x$, if the outcome is $+1$ the quantum state is $|+\rangle = \frac{1}{\sqrt{2}}\left(|0\rangle + |1\rangle\right)$, and if the outcome is $-1$ the quantum state is $|-\rangle = \frac{1}{\sqrt{2}}\left(|0\rangle - |1\rangle\right)$. Therefore, the quantum state after each measurement is always one of the four quantum states $|0\rangle$, $|1\rangle$, $|+\rangle$, and $|-\rangle$. The quantum state after the measurement at time $t$ is the quantum state before the measurement at $t+1$. The process is repeated infinitely many times. If the measurements are randomly chosen, the quantum state has probability $\frac{1}{2}$ to change and probability $\frac{1}{2}$ not to change.

We consider type I interpretations satisfying the following assumptions: (i) The choice of which measurement is performed can be made randomly and independently of the system under observation, (ii) the system has limited memory, and (iii) Landauer's erasure principle holds \cite{Landauer61}. In these interpretations, at any time $t$ the quantum probabilities are determined by intrinsic properties of the system and, according to assumption (i), these intrinsic properties change depending on what measurements are performed. Assumption (ii) implies that the system cannot have stored the values of the intrinsic properties for all possible sequences of measurements that the observer can perform. This implies that the system has to generate new values and store them in its memory. For that reason, the system needs to erase part of the previously existing information. Landauer's principle states that the erasure of one bit of information from the information-bearing degrees of freedom of a system must be accompanied by a corresponding entropy increase in the non-information-bearing degrees of freedom of the system, therefore causing dissipation of at least $k T\ln 2$ units of heat, where $k$ is the Boltzmann constant and $T$ is the temperature of the system. Landauer's principle has been verified in actual experiments \cite{BAPCDL12,JGB14} and is considered valid in the quantum domain \cite{HSAL11,RW14}. Therefore, whenever the temperature is not zero, assumption (iii) implies that the system should dissipate at least an amount of heat proportional to the information erased. Since in type I interpretations satisfying assumptions (i)--(iii) the system under observation will be represented by a finite-state machine with more than one state, we can invoke the third law of thermodynamics (i.e., it is impossible by any procedure, no matter how idealized, to reduce the temperature of any system to zero temperature in a finite number of finite operations) to assume that the erasure of information in the system occurs at non zero temperature.

%%%%%%%%%%%%%%%%%%%%%%%%%%%%%%%%%%%%%%%%%%%%%%%%%%%%%%%%%%%%%%%%%%%

\section{Tools}
\label{Sec:3}

%%%%%%%%%%%%%%%%%%%%%%%%%%%%%%%%%%%%%%%%%%%%%%%%%%%%%%%%%%%%%%%%%%%

In type I interpretations the system generates outcomes whose probabilities (including the case of probability equal to 1) are determined only by intrinsic properties of the system, that is, by the state of the memory of the system prior to the measurement. To calculate the minimum information that the system must erase per measurement, a key observation is that our ideal experiment is an example of a stochastic input-output process that can be analyzed in information-theoretic terms.

A stochastic process $\overleftrightarrow{{\cal Y}}$ is a one-dimensional chain $\ldots,Y_{-2},Y_{-1},Y_0,Y_1,Y_2,\ldots$ of discrete random variables $\{Y_t\}_{t \in \mathbb{Z}}$ that take values $\{y_t\}_{t \in \mathbb{Z}}$ over a finite or countably infinite alphabet ${\cal Y}$. An input-output process $\overleftrightarrow{Y}|\overleftrightarrow{X}$ with input alphabet ${\cal X}$ and output alphabet ${\cal Y}$ is a collection of stochastic processes $\overleftrightarrow{Y}|\overleftrightarrow{X} \equiv \{ \overleftrightarrow{Y} | \overleftrightarrow{x} \}_{\overleftrightarrow{x} \in \overleftrightarrow{{\cal X}}}$, where each such process $\overleftrightarrow{Y}|\overleftrightarrow{x}$ corresponds to all possible output sequences $\overleftrightarrow{Y}$ given a particular bi-infinite input sequence $\overleftrightarrow{x}$. It can be represented as a finite-state automaton or, equivalently, as a hidden Markov process. In our experiment, $x_t$ is the observable measured at time $t$ and $y_t$ the corresponding outcome. By $\overleftarrow{X}$ we denote the chain of previous measurements, $\ldots,X_{t-2},X_{t-1}$, by $\overrightarrow{X}$ we denote $X_t,X_{t+1},\ldots$, and by $\overleftrightarrow{X}$ we denote the chain $\ldots,X_{t-1},X_{t},X_{t+1},\ldots$. Similarly, $\overleftarrow{Y}$, $\overrightarrow{Y}$, and $\overleftrightarrow{Y}$ denote the past, future, and all outcomes, respectively, while $\overleftarrow{Z}$, $\overrightarrow{Z}$, and $\overleftrightarrow{Z}$ denote the past, future, and all pairs of measurements and outcomes. For deriving physical consequences we have to consider the minimal and optimal representation of this process.

The fact that our experiment is an input-output process implies \cite{BC14} that there exists a unique minimal and optimal predictor of the process, i.e., a unique finite-state machine with minimal entropy over the state probability distribution and maximal mutual information with the process's future output given the process's input-output past and the process's future input. This machine is called the process's $\varepsilon$-transducer \cite{BC14} and is the extension of the so-called $\varepsilon$-machines \cite{CY89, SC01}. An $\varepsilon$-transducer of an input-output process is a tuple $({\cal X}, {\cal Y}, {\cal S}, {\cal T})$ consisting of the process's input and output alphabets ${\cal X}$ and ${\cal Y}$, the set of causal states ${\cal S}$, and the set of corresponding conditional transition probabilities ${\cal T}$. The causal states $s_{t-1} \in {\cal S}$ are the equivalence classes in which the set of input-output pasts $\overleftarrow{{\cal Z}}$ can be partitioned in such a way that two input-output pasts $\overleftarrow{z}$ and $\overleftarrow{z}'$ are equivalent if and only if the probabilities $P(\overrightarrow{Y}|\overrightarrow{X},\overleftarrow{Z}=\overleftarrow{z})$ and $P(\overrightarrow{Y}|\overrightarrow{X},\overleftarrow{Z}=\overleftarrow{z}')$ are equal. The causal states are a so-called sufficient statistic of the process. They store all the information about the past needed to predict the output and as little as possible of the remaining information overhead contained in the past. The Shannon entropy over
the stationary distribution of the causal states $H({\cal S})$ is the so-called statistical complexity and represents the minimum internal entropy needed to be stored to optimally compute future measurement outcomes (this quantity generally depends on how our measurements $\overleftrightarrow{X}$ are selected; here we assume each $X_t$ is selected from a uniform probability distribution). The set of conditional transition probabilities ${\cal T} \equiv \{P(S_{t+1} = s_j,Y_t=y|S_t=s_i,X_t=x)\}$ governs the evolution.

The fact that the $\varepsilon$-transducer is also the machine producing minimum heat can be proven as follows. The average information that must be erased per measurement is the information contained in the causal state previous to the measurement, $S_{t-1}$, that is not contained in the causal state after the measurement, $S_{t}$ \cite{WGRV12}. This is equal to the conditional entropy (or uncertainty) of $S_{t-1}$ given $S_t$, $X_t$, and $Y_t$,
\begin{equation}
 \label{eq:1}
 I_{\rm erased} = H (S_{t-1} | X_t, Y_t, S_t).
\end{equation}
For the machine reproducing the outcomes of our experiment, knowledge of
the current state gives complete information about the last outcome, so 
$H(Y_t | X_t, S_t) = 0$. Then the properties of conditional entropy give
\begin{equation}
 \label{eq:2}
 I_{\rm erased} = H(S_t|X_t,S_{t-1})-H(S_t|X_t)+H(S_{t-1}|X_t).
\end{equation}
We note that $H(S_{t-1}|X_t)=H(S_{t-1})=H(S_t)$ because measurements are
chosen at random and $\varepsilon$-transducers are stationary, so for our
experiment the amount of information erased is
\begin{equation}
 \label{eq:3}
 I_{\rm erased} = H(S_t|X_t,S_{t-1})+I(S_t:X_t),
\end{equation}
where the mutual information $I(S_t:X_t)=H(S_t)-H(S_t|X_t)$. States $R_t$ of another partition $\cal R$ of input-output pasts $\overleftarrow{Z}$ that are as predictive as the causal states \cite{SC01} obey $H(R_t|X_t,R_{t-1})\ge{}H(S_t|X_t,S_{t-1})$ and the data-processing inequality (see, e.g., Ref.~ \cite{CT06}) guarantees that $I(S_t:X_t)\le I(R_t:X_t)$. It follows that the quantity $I_{\rm erased}$ is minimal for an $\varepsilon$-transducer.

In addition, the $\varepsilon$-transducer of our experiment has a particular property, namely, that there is a one-to-one correspondence between causal states $s_t$ and quantum states $|\psi_{t}\rangle \in \Psi$. All input-output pasts in a given causal state $s_t$ give the same probabilities for all $\overrightarrow{X}$, so the quantum states given by the input-output pasts must be equal. Conversely, a given quantum state provides probabilities for all $\overrightarrow{X}$ that correspond to one causal state of the $\varepsilon$-transducer. As a consequence, the $\varepsilon$-transducer associated with our experiment has four causal states that we will denote $s_0$, $s_1$, $s_+$, and $s_-$, corresponding to the four quantum states $|0\rangle$, $|1\rangle$, $|+\rangle$, and $|-\rangle$.

%%%%%%%%%%%%%%%%%%%%%%%%%%%%%%%%%%%%%%%%%%%%%%%%%%%%%%%%%%%%%%%%%%%

\section{Results}
\label{Sec:4}

%%%%%%%%%%%%%%%%%%%%%%%%%%%%%%%%%%%%%%%%%%%%%%%%%%%%%%%%%%%%%%%%%%%

{\em Result~1.} For the experiment in which a qubit is submitted to sequential measurements randomly chosen from ${\cal X}=\{\sigma_x,\sigma_z\}$ at temperature $T$, under assumptions (i)--(iii), any interpretation of quantum theory in which probabilities are determined by intrinsic properties of the system predicts that, on average, the system should dissipate at least $\frac{3}{2} k T \ln 2$ units of heat per measurement.

%%%%%%%%%%%%%%%%%%%%%%%%%%%%%%%%%%%%%%%%%%%%%%%%%%%%%%%%%%%%%%%%%%%

{\em Proof:} Since, in our experiment, the stationary distribution of the causal states is uniform and the observer randomly chooses the measurement, we can calculate $I_{\rm erased}$ by using a particular causal state, e.g., $S_t=s_0$, and a particular measurement that lead to it, e.g., $X_t=\sigma_z$. Using Eq.~(\ref{eq:1}) and the fact that $H(Y_t | S_t)=0$ we obtain
\begin{equation}
\label{eq:4}
\begin{split}
 &I_{\rm erased} = H (S_{t-1} | \sigma_z, s_0) \\
 &= - \sum_{s_j \in {\cal S}} P(S_{t-1}=s_j | \sigma_z, s_0) \log_2 P(S_{t-1}=s_j | \sigma_z, s_0),
\end{split}
\end{equation}
where logarithms are to base two, $0 \log_2 0$ is taken as zero, and $P(S_{t-1}=s_j | \sigma_z, s_0)$ is the conditional probability that the causal state at time step $t-1$ was $s_j$, given a measurement choice $\sigma_z$ and subsequent transition to causal state $s_0$ at time step $t$. There are only three possible causal states at time $t$: $s_0$, $s_+$, and $s_-$, whose respective conditional probabilities $P(S_{t-1}=s_j | \sigma_z, s_0)$ are $\frac{1}{2}$, $\frac{1}{4}$, and $\frac{1}{4}$. Therefore,
\begin{equation}
\label{eq:5}
 I_{\rm erased}=-\frac{1}{2} \log_2 \frac{1}{2}-2 \frac{1}{4} \log_2 \frac{1}{4}=\frac{3}{2}\;{\rm bits}.
\end{equation}
Applying Landauer's principle, we obtain the corresponding lower bound to the average heat per measurement dissipated by the system. \hfill \endproof

In contrast, if probabilities are not determined by intrinsic properties of the system, then measurement outcomes are created randomly when the observables are measured, without any need to overwrite information in the system and therefore without the system dissipating heat due to Landauer's principle. Landauer's principle also applies to the measurement apparatus; we observe that there is no difference between type I interpretations satisfying assumptions (i)--(iii) and type II interpretations regarding the heat dissipated by the measurement apparatus. What is different is that the first type of interpretations requires, in addition, a finite-state machine representing the system under observation (which is absent in type II interpretations), which produces an {\em extra} amount of heat due to Laudauer's principle (which is not produced in type II interpretations).

Result~1 assumes that the observer can only choose between two measurements. What if the observer chooses measurements from a larger set?

%%%%%%%%%%%%%%%%%%%%%%%%%%%%%%%%%%%%%%%%%%%%%%%%%%%%%%%%%%%%%%%%%%%

{\em Result~2.} For the experiment in which a qubit is submitted to sequential measurements randomly chosen from
\begin{equation}
 \label{eq:6}
 {\cal X}(n)=\Big\{\cos\left(\tfrac{\pi k}{2^n}\right)\sigma_z+\sin\left(\tfrac{\pi k}{2^n}\right)\sigma_x,k=0,\ldots,2^n-1\Big\},
\end{equation}
under assumptions (i)--(iii), any interpretation of quantum theory in which probabilities are determined by intrinsic properties of the system predicts that, on average, the system should dissipate an amount of heat per measurement that tends to infinity linearly with $n$.

%%%%%%%%%%%%%%%%%%%%%%%%%%%%%%%%%%%%%%%%%%%%%%%%%%%%%%%%%%%%%%%%%%%

{\em Proof:} The sets ${\cal X}(n)$ and
{\small
\begin{equation}
 \label{eq:7}
 \Psi(n)=\Big\{\cos\left(\tfrac{\pi j}{2^{n+1}}\right)|0\rangle
 +\sin\left(\tfrac{\pi j}{2^{n+1}}\right)|1\rangle,
 j=0,\ldots,2^{n+1}-1\Big\}
\end{equation}}
have, for any $n \in \mathbb{N}$, the same properties we exploited for the proof of result~1. Indeed, ${\cal X}(1)={\cal X}$, $\Psi(1)=\Psi$, and $s_0$ is associated with $|0\rangle$ for all $n$. Therefore, we can follow the same strategy as in the proof of result~1. The number of pure quantum states attained by the system now grows exponentially with $n$ and, consequently, the number of causal states in the associated $\varepsilon$-transducer also grows exponentially with $n$ and the needed conditional probability is
\begin{equation}
 \label{eq:8}
 P(S_{t-1}=s_j|\sigma_z,s_0)=\frac{\cos^2\left(\frac{\pi j}{2^{n+1}}\right)}{\sum_{l=0}^{2^{n+1}-1}\cos^2\left(\frac{\pi l}{2^{n+1}}\right)}=\frac{\cos^2\left(\frac{\pi j}{2^{n+1}}\right)}{2^n}.
\end{equation}
Inserting this probability in the erased information, and using that $\cos^2\phi<1$ when $0<\phi<\pi$ inside the logarithm, we obtain
\begin{equation}
 \label{eq:9}
 I_{\rm erased}(n)=-\sum_{j=0}^{2^{n+1}-1}\frac{\cos^2\left(\frac{\pi j}{2^{n+1}}\right)}{2^n}
\log_2 \frac{\cos^2\left(\frac{\pi
j}{2^{n+1}}\right)}{2^n}>n.
\end{equation}
Therefore, when the temperature of the system is not zero, the lower bound of the heat dissipated by the system from Landauer's principle tends to infinity at least linearly with~$n$. \hfill \endproof

%%%%%%%%%%%%%%%%%%%%%%%%%%%%%%%%%%%%%%%%%%%%%%%%%%%%%%%%%%%%%%%%%%%

\section{Conclusions}
\label{Sec:5}

%%%%%%%%%%%%%%%%%%%%%%%%%%%%%%%%%%%%%%%%%%%%%%%%%%%%%%%%%%%%%%%%%%%

Here we have shown that, for those interpretations of quantum theory in which quantum probabilities are determined by intrinsic properties of the observed system, assumptions (i)--(iii) lead to the prediction that the observed system must dissipate an unbounded amount of heat in each measurement due to Landauer's principle. This conclusion is independent of how the observed system is defined. The only relevant assumption about the observed system is that it is finite.

As a reaction to this result one has the following options.

(a) One could abandon the assumption that quantum probabilities are determined by intrinsic properties of the observed system.

(b) One could abandon assumption (i), which would mean that the decision of which measurement is performed on the system cannot be made randomly and independently of the system.

(c) One could abandon assumption (ii), which would mean that, at any given time, a finite system contains infinite information.

Abandoning both assumptions (i) and (ii) would allow for superdeterminism. Trying to exclude superdeterminism on scientific grounds seems ``beyond the power of logic'' \cite{Wheeler83}.

(d) One could abandon assumption (iii), which would mean that Landauer's principle does not apply at the level of type I interpretations. This would mean that the intrinsic properties of the observed system behind quantum probabilities satisfy different physical laws than other intrinsic properties.

(e) One could assume that the erasure of information in the system can occur at zero temperature. This would violate the third law of thermodynamics.

(f) One could assume that the set of possible measurements and states is actually discretized. Toy theories with this property have been shown to retain fundamental features of quantum theory \cite{SW12,HOST13}. This solution would imply that quantum theory is only a continuous idealization of a deeper discrete theory.

(g) One could assume that the precision of the measurements is fundamentally limited.

Nevertheless, even in cases (f) and (g), the observed system should dissipate a potentially observable amount of heat that would depend on the size of the system, the length of the discretization, and/or the limits in the precision of the measurements. Bounds on these quantities may be established through precision experiments measuring this hypothetical heat. Such an experiment can, in principle, be conducted using two ions confined in separate wells: The target qubit is encoded in the internal state of one of the ions, while the other ion is used as an ancillary qubit for readout \cite{BGLTHJHLW12,KRSSP14}.

The aim of this article has been to introduce a fresh perspective, thermodynamics, in one of the longest-standing debates in physics, the interpretation of quantum theory. Our aim has not been excluding any specific interpretation, but to point out the consequences of some natural assumptions in those interpretations in which quantum probabilities are determined by intrinsic properties of the observed system. In this sense, one can draw a parallel with Bell's theorem \cite{Bell64}. Bell's theorem does not exclude Bohmian mechanics. It just shows that the assumption of local realism is incompatible with quantum theory and, as a consequence, points out that any realistic interpretation of quantum theory cannot be local. Similarly, our result does not exclude Bohmian mechanics or the many worlds interpretation, since in both cases assumption (ii) is not satisfied (in Bohmian mechanics because the observed system includes an underlying continuous field and in the many worlds interpretation because the system itself splits in each measurement), but draws attention to the fact that systems with unlimited memory are needed in any interpretation in which quantum probabilities are determined by intrinsic properties of the observed system.

%%%%%%%%%%%%%%%%%%%%%%%%%%%%%%%%%%%%%%%%%%%%%%%%%%%%%%%%%%%%%%%%%
% Acknowledgements
%%%%%%%%%%%%%%%%%%%%%%%%%%%%%%%%%%%%%%%%%%%%%%%%%%%%%%%%%%%%%%%%%

\begin{acknowledgments}
This work was supported by the FQXi large grant project ``The Nature of Information in Sequential Quantum Measurements'', Projects No.\ FIS2011-29400 and No.\ FIS2014-60843-P (MINECO, Spain) with FEDER funds, and the DFG. 
M.G. was supported by the John Templeton Foundation Grant No.\ 54914 ``Occam's Quantum Mechanical Razor: Can Quantum Theory admit the Simplest Understanding of Reality?'' and the Singapore National Research Foundation under NRF Award No.\ NRF–NRFF2016–02.
O.G. acknowledges funding from the ERC (Consolidator Grant No.\ 683107/TempoQ). 
K.W. acknowledges EPSRC support via Grant No.\ EP/E5012141. 
We thank D.\ Z.\ Albert, J.\ Anders, M.\ Ara\'ujo, L.\ E.\ Ballentine, H.\ R.\ Brown, \v{C}.\ Brukner, J.\ Bub, C.\ Budroni, J.\ I.\ Cirac, D.\ Dieks, C.\ A.\ Fuchs, R.\ B.\ Griffiths, M.\ Kleinmann, M.\ Leifer, O.\ Lombardi, L.\ Maccone, N.\ D.\ Mermin, R.\ Schack, J.\ Thompson, C.\ Timpson, C.\ Schmiegelow, A.\ G.\ White, and C.\ Wunderlich for conversations on different aspects of the problem addressed here.
\end{acknowledgments}

%%%%%%%%%%%%%%%%%%%%%%%%%%%%%%%%%%%%%%%%%%%%%%%%%%%%%%%%%%%%%%%%%%%
%
% Foundational Questions Institute (FQXi) FundRef ID http://dx.doi.org/10.13039/100009566
% FEDER: Fonds europ\'een de d\'eveloppement \'economique et r\'egional
% MINECO: Ministerio de Econom\'{\i}a y Competitividad FundRef ID http://dx.doi.org/10.13039/501100001862
% European Research Council (ERC) http://dx.doi.org/10.13039/501100000781
%
%%%%%%%%%%%%%%%%%%%%%%%%%%%%%%%%%%%%%%%%%%%%%%%%%%%%%%%%%%%%%%%%%%%

\end{document}